# Angriffserkennung für industrielle Netzwerke innerhalb des Projektes IUNO

Simon DUQUE ANTON, Daniel FRAUNHOLZ, Hans Dieter SCHOTTEN
Deutsches Forschungszentrum für Künstliche Intelligenz, Kaiserslautern, Deutschland,
{Vorname}.{Nachname}@dfki.de

## Kurzfassung

Die zunehmende Vernetzung in industriellen Netzwerken ist eines der zentralen aktuellen Themen. Sie wird sowohl durch Unternehmen, als auch durch Forschungsinstitute adressiert. Um den Schritt zur vierten industriellen Revolution hin zu vollziehen, ist eine durchgehende Konnektivität von Produktionseinrichtungen unerlässlich. Aufgrund dieser zunehmenden Vernetzung entstehen jedoch auch neue Angriffsvektoren für das industrielle Umfeld. Im Nationalen Referenzprojekt zur IT-Sicherheit in der Industrie 4.0 (IUNO) werden die entstehenden Risiken und Bedrohungen aufgegriffen und Schutzmaßnahmen entwickelt. Diese Schutzmaßnahmen sind geeignet, auch von kleinen und mittelständischen Unternehmen ohne großen Aufwand umgesetzt zu werden, da diese sich aufgrund geringerer personeller und finanzieller Mittel in einer besonderen Bedrohungssituation befinden. Die entwickelten Schutzmaßnamen werden anhand von vier verschiedenen Anwendungsfällen prototypisch umgesetzt. Ein weiteres Augenmerk liegt auf Ergebnissen und Forschungsfeldern des DFKI, die im Rahmen des Projektes IUNO entstanden sind. Neben der Vorstellung von IUNO wird im Rahmen dieser Arbeit eine Methode für netzwerkbasierte verteilte Datenaggregation vorgestellt, die zur Anomalieerkennung notwendig ist.

## 1    Einleitung

Momentan ist die industrielle Landschaft, insbesondere in Deutschland, im Wandel begriffen. Die Nachfrage an Vernetzungen, sowohl innerhalb einer Produktionseinrichtung, als auch fabrikübergreifend steigt stetig. Dieser Wandel steht in engem Zusammenhang mit Flexibilität, die von den Herstellern immer nachdrücklicher gefordert wird. Kundenindividuelle Produktion ist ein aktuelles Thema, genauso wie Fernwartung, sichere Daten und die visuelle Darstellung von Sicherheitszuständen. Diese erhöhte Flexibilität, gepaart mit erweiterter Vernetzung, birgt jedoch zahlreiche neue Angriffsvektoren für die industriellen Netzwerke. Aktuelle Netzwerksicherheitsmaßnahmen wie Antivirenprogramme, Firewalls und statische Angriffserkennung und -prävention können den wachsenden Anforderungen an die Sicherheit der modernen Industrie nicht gerecht werden.

Um Stellung von Deutschland als Wirtschaftsstandort auch im Rahmen dieser Entwicklungen zu sichern, wurde das Nationalen Referenzprojekt zur IT-Sicherheit in der Industrie 4.0 (IUNO) initiiert. Im Rahmen des Projektes IUNO werden zahlreiche Themen adressiert. Sichere (Funk-)Kommunikation bildet die Grundlage einer sicheren, vernetzten Infrastruktur. Angriffe, von innen und von außen, werden anhand fortschrittlicher Sicherheitsanalysen im Vorfeld klassifiziert. In der Anwendung werden sie von neuartigen Algorithmen und Werkzeugen erkannt, sodass Gefahren schnell gebannt werden können. Auch eine übersichtliche und aussagekräftige Darstellung der aktuellen Bedrohungslage gehört zu einem umfassenden Sicherheitskonzept. Der Schutz von geistigem Eigentum (englisch: Intellectual Property, IP) gewinnt zunehmen an Bedeutung, da die Wertschöpfung vieler Firmen nicht länger in dem eigentlichen physischen Produkt besteht, sondern im Wissen um die Verarbeitung und Herstellung dieses Produktes. Die entwickelten Technologien und Werkzeuge werden im Rahmen von IUNO anhand von Demonstratoren anschaulich dargestellt. Insgesamt werden vier Demonstratoren entwickelt: Eine beispielhafte Umsetzung von kundenindividueller Produktion in der Fertigung, ein Technologiedatenmarktplatz, bei dem Verwendung der Technologiedaten durch den Anbieter limitiert wird, einem visuellen IT-Sicherheitsleitstand und einem neuartigen Konzept zur flexiblen und sicheren Fernwartung.

Die Arbeit ist aufgebaut wie folgt: Zunächst wird in Kapitel 2 ein Überblick über das Forschungsprojekt IUNO gegeben. Anschließend wird in Kapitel 3 ein aktuelles Forschungsthema innerhalb von IUNO vorgestellt. In Kapitel 4 wird ein Fazit gezogen und weitere Schritte umrissen

## 2    Das Forschungsprojekt IUNO

In diesem Kapitel wird das Nationalen Referenzprojekt zur IT-Sicherheit in der Industrie 4.0 vorgestellt. Zunächst werden in einer Übersicht das Projekt und seine Ziele erläutert. Anschließend werden ausgewählte Methoden, die vom Deutschen Forschungszentrum für Künstliche Intelligenz (DFKI) im Kontext von IUNO entwickelt wurden, genannt. Die von IUNO adressierten Anwendungsfälle werden abschließend dargelegt.

### 2.1    Übersicht

Ziel des Nationalen Referenzprojekts zur IT-Sicherheit in der Industrie 4.0 ist es, die Lücke zwischen Anforderung und Angebot der aktuellen IT-Sicherheitstechnologie zu schließen. Ein besonderer Fokus liegt auf der Einbindung kleiner und mittelständischer Unternehmen (KMU). Diese sollen in die Lage versetzt werden, die entwickelten Lösungen und Werkzeuge ohne großen Aufwand in ihr Repertoire zur Cyberabwehr zu übernehmen. Dies ist vor allem deshalb wichtig, da KMU oft nicht über große IT-Sicherheitsabteilungen verfügen. Weiterhin ist gerade am Wirtschaftsstandort Deutschland eine Prävalenz von KMU im industriellen Umfeld zu spüren. Ein Leitfaden zur Industrie 4.0 findet sich in [1].

## 2.2 Methoden des DFKI

Einige der im Rahmen von IUNO bereits entwickelten und publizierten Methoden werden in diesem Kapitel vorgestellt. Neben grundlegenden Arbeiten wie Klassifizierungen von Angriffen [2] wurde ein großer Beitrag im Bereich der Physical Layer Security geleistet. In diesem Forschungsbereich werden die Eigenschaften der Schicht 1 des Open Systems Interconnection (OSI)-Modells für die Erzeugung von Sicherheitsmerkmalen verwendet. Auch im Bereich dezeptiver Systeme wurde geforscht. Diese dienen dazu, Angreifer mit scheinbar verwundbaren Systemen zu locken und ihr Verhalten zu erfassen. Schließlich wurde ein Modell zur Aggregation von Daten entwickelt.

### 2.2.1 Physical Layer Security

Innerhalb des Projektes wurde eine vorab konzeptionierte Methode für eine verteile Schlüsselgenerierung und –synchronisierung [3 bis 5] zu einem einsatzfähigen Demonstrator entwickelt. Diese Methode benötigt keinerlei vorab verteilte Information zur Generierung eines kryptographischen Schlüssels. Vielmehr wird ausgehend von den Eigenschaften eines drahtlosen Kanals ein Schlüssel erzeugt, der durch fehlerkorrigierende Codes abgeglichen und zur Erzeugung eines kryptographischen Sitzungsschlüssels verwendet wird. Die Eigenschaften eines Funkkanals sehen für die Kommunikationspartner annährend gleich aus, eine Eigenschaft, die als Reziprozität bezeichnet wird. Ein Angreifer, der auf dem drahtlosen Kanal lauscht, wird unweigerlich andere Kanaleigenschaften sehen, sofern er physisch entfernt positioniert ist. Als geeignete Eigenschaft zur Erzeugung eines Schlüssels hat sich insbesondere der Returned Signal Strength Indicator (RSSI) erwiesen.

Insbesondere bei der Verwendung vieler eingebetteter Systeme im industriellen Umfeld ergeben sich zahlreiche Vorteile durch derartige dezentrale Schlüsselerzeugung und -management. Rechenintensive Operationen wie bei klassischer Kryptographie werden nicht benötigt. Eine a priori Ausbringung der Schlüssel ist nicht notwendig. Es besteht forward- und backward-secrecy, bei der Kompromittierung eines Schlüssels sind vergangene und zukünftige Kommunikationen nicht betroffen. Dies gilt bei regelmäßiger Schlüssel-Neugenerierung, was in der Praxis trivial und ratsam ist.

Dieses Konzept unterliegt einigem Forschungsinteresse und wird in aktuellen Arbeiten weiterverfolgt [6]. Auch Methoden des maschinellen Lernens werden eingesetzt, um die Qualität von Physical Layer Security zu erhöhen [7].

### 2.2.2 Dezeptive Verteidigungsmaßnahmen

Neure Studien belegen ein gesteigertes Interesse von Kriminellen gegenüber industrieller Infrastruktur [8, 9]. Dabei sind besonders Angriffe von Botnets besonders interessant, da diese eine große Verbreitung aufweisen und jedes mit dem Internet verbundene System bedroht ist [10]. Am DFKI wurden Methoden untersucht, um Angriffe auf industrielle Netze auch nach erfolgreicher Kompromittierung zu erkennen. Dazu wurden sogenannte dezeptive Abwehrmaßnahmen z.B. Honeypots entwickelt und eingesetzt. Kern dieser Maßnahmen ist ein gezieltes täuschen von Eindringlingen, um Schäden von Produktionssystem zu mitigieren und gleichzeitig Informationen über den Angreifer und die verwendeten Angriffsvektoren zu sammeln. Mit den gesammelten Daten können weitere Angriffsversuche unterbunden werden und unter Umständen auch eine Identifikation des Angreifers möglich gemacht werden. In einem ersten Schritt wurden vorhandene Lösungen evaluiert und gängige Angriffsmethoden analysiert [11], dazu wurde ein Honeynet eingesetzt. Honeynets eignen sich ausgezeichnet zum Erfassen von Trends und Statistiken zu aktuellen Entwicklungen in der Bedrohungslandschaft [12]. Im weiteren Verlauf wurden Verteilkonzepte für dynamische Netzwerke [13], grundlegende Gestaltungsprinzipien [14] und maschinelle Lernansätze zur forensischen Analyse der aufgenommen Daten [15] untersucht.

### 2.2.3 Aggregationsmodell

Ein großes Ziel des Projektes IUNO ist die zuverlässige Erkennung von Anomalien in einem Netzwerk. Hierzu gibt es bereits zahlreiche vorarbeiten, gut zusammengefasst etwa in [16, 17]. Zur Erkennung von Angriffen und Anomalien in Netzwerken ist es zunächst notwendig, Daten zu sammeln. Diese Daten stammen aus unterschiedlichen Quellen und liegen in unterschiedlichen Formaten vor. Nach der Sammlung müssen die Daten zusammengeführt und inhaltlich reduziert werden. Dieser Vorgang wird im Forschungsfeld des Complex Event Processing (CEP) behandelt [18, 19]. Oftmals sind für bestimmte Anwendungsfälle nicht alle gesammelten Informationen notwendig. Durch das Zusammenführen der Daten ist ein umfassenderes Bild des Netzwerk- oder Systemzustandes ersichtlich, da mehrere Perspektiven eines Ereignisses in Betracht gezogen werden [20]. Diese Datensammlung und –aggregation dient als Grundlage für die Anomalieerkennung, die stark abhängig ist von der Qualität der verwendeten Daten ist.

## 2.3 Anwendungsfälle

Die in IUNO analysierten Anwendungsfälle fokussieren jeweils eins der vier Kernthemen von Industrie 4.0: sichere Prozesse, sichere Daten, sichere Dienste und sichere Vernetzung.

### 2.3.1 Kundenindividuelle Produktion

Der Trend der industriellen Fertigung divergiert von Massenproduktion hin zu Fertigungsserien, die Stück für Stück unterschiedlich sind, also einer Losgröße von eins. Hierbei müssen die durchzuführenden Bearbeitungen für jedes Werkstück individuell erkannt und ausgeführt werden. Zu diesem Zweck werden die Werkstücke mit Intelligenz versehen, also mit Informationen, wie mit ihnen zu verfahren ist. Hierzu gibt es bereits Methoden [21], die jedoch auch Schwächen haben. Barcodes können verschmutzen und unleserlich werden oder beim Bearbeitungsprozess in Mitleidenschaft gezogen werden. Radio Frequency Identification (RFID)-Tags sind im Vergleich zu Barcodes teuer, außerdem müssen sie derart im Werkstück platziert werden, dass sie leicht ausgelesen werden können, jedoch durch die Bearbeitung nicht beschädigt oder entfernt werden, etwa bei einem Sägevorgang.

### 2.3.2 Technologiedaten-Markplatz

Wie in Kapitel 1 bereits erwähnt kommt dem IP eine zunehmende Relevanz bei der Wertschöpfung zu. Das eigentliche Produkt verliert dagegen an Bedeutung, vielmehr macht das technische Know-How um die Verarbeitung die Stellung eines Unternehmens aus. Dieses Wissen wird in steigendem Maße monetarisiert, etwa durch den Verkauf von Schaltplänen [22]. Zukünftig soll dieses Geschäftsfeld auch auf die verarbeitende Industrie ausgeweitet und die Infrastruktur für eine sichere Lizensierung geschaffen werden. Es ist kritischer Teil dieser Aufgabe, dass die Nutzung überlassener Daten zuverlässig nur in dem Rahmen passiert, wie die Lizensierung es erlaubt. Dadurch wird Diebstahl geistigen Eigentums und Produktpiraterie entgegengewirkt. Diese Probleme werden im Rahmen von IUNO durch einen Technologiedaten-Markplatz gelöst. Auf diesem Markplatz treffen Technologiedaten-Anbieter und –Konsumenten aufeinander. Sie einigen sich auf Konditionen, anschließend verwendet der Konsument die Daten entsprechend der Übereinkunft. Hierbei wird durch den Marktplatz sichergestellt, dass eine abweichende Verwendung der überlassenen Daten ausgeschlossen ist. Zu diesem Thema wurden von der TU Darmstadt Konzepte entwickelt [25].

### 2.3.3 Fernwartung / Trusted Partner

Die Nichtverfügbarkeit von Produktionseinrichtungen ist ein großer Kostentreiber für produzierende Unternehmen. Um störungs- und wartungsbedingte Standzeiten zu minimieren ist Fernwartung eine zukunftsfähige Möglichkeit. Hierbei muss der Fernwartungstechniker nicht mehr vor Ort sein, sondern greift per internetbasiertem Fernzugriff auf die entsprechende Maschine zu. Dadurch wird die Reaktionszeit erhöht, da kürzere Zeiten zwischen Meldung und Behebung einer Störung entstehen. Außerdem ist die Wartung unabhängig vom Standort der Maschine. Jedoch wird auch ein großes Vertrauen in die Fernwartungsinfrastruktur vorausgesetzt. Eine Wartung darf nur stattfinden, sofern ein Wartungsfall vorliegt, andernfalls darf kein Zugriff auf Produktionseinrichtungen zugelassen werden. Auch wenn ein Wartungsfall vorliegt, darf der Zugriff nur durch eindeutig identifizierte, ausgewählte Personen erfolgen. Die Verbindung muss dabei zusätzlich stets vertraulich und unverändert geschehen. Dies wird durch eine zentrale Cloud-Plattform sichergestellt. Innerhalb dieser Cloud-Plattform werden die Wartungstickets und kryptographisches Material verwaltet. Weiterhin geschieht das Verbindungsmanagement durch diese Plattform. Für einen Maschinenbetreiber ist es nun möglich, im Wartungsfall ein Ticket zu erstellen, welches von einem validierten Fernwartungsanbieter akzeptiert wird. Nun wird eine ad-hoc-Verbindung aufgebaut, im Rahmen derer die Fernwartung geschieht.

### 2.3.4 Visueller Security-Leitstand

Um die Sicherheitslage eines Produktionsnetzwerkes einschätzen zu können, ist es notwendig, Daten zu sammeln, zusammenzuführen, zu bewerten und anschließend graphisch leicht verständlich darzustellen. Durch Einbringung schadhaften Netzwerkverkehrs in Produktionsnetze besteht ein enormes Schadenspotential. Daher ist es unerlässlich, Angriffsversuche frühzeitig zu erkennen und zu mitigieren. Zunächst wird hierzu Konzept entwickelt und implementiert, welches eine Erfassung des relevanten Netzwerkverkehrs der Produktionsanlage ermöglicht. Zusätzlich werden geeignete Algorithmen entwickelt, um die gesammelten Daten in ein handhabbares Datenformat zu aggregieren. Die Aggregation soll dabei eine möglichst große Komprimierung bei minimalem Qualitätsverlust ermöglichen. Auch die zur Auswertung der gesammelten Daten nötigen Algorithmen werden in IUNO entwickelt, implementiert und getestet. Diese Algorithmen sind geeignet, den Sicherheitszustand des Produktionsnetzwerkes zu beschreiben und wiederzugeben. Abschließend werden die gewonnenen Informationen graphisch ansprechend und verständlich wiedergegeben, sodass sie leicht zu interpretieren sind. Ansätze der Technischen Universität München etwa umfassen die Entwicklung eines Frameworks zur Analyse maliziöser Dateien sowie die Untersuchung von verdächtigen Anwendungsdateien [23, 24].

## 2.4 Anwendung der Methoden

Die in Kapitel 2.2 beschriebenen Methoden lassen sich in die Lösung der Anwendungsfälle integrieren. Aufgrund ihrer Modularität und der Tatsache, dass sie als eigenständige Konzepte entwickelt wurden, können sie in unterschiedlichen Kontexten sinnvoll verwendet werden. Eine Matrix, welche die Anwendbarkeit der Methoden in den Lösungsansätzen darstellt, ist in Abbildung 1 gegeben.

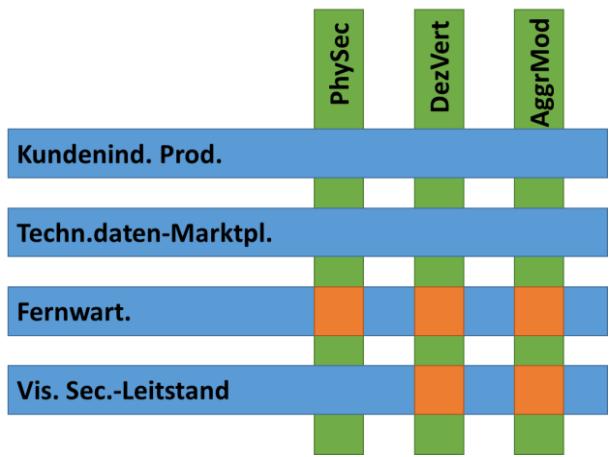

**Abbildung 1: Vergleich Methoden und Lösungsansätze in IUNO**

# 3 Schwerpunkt: Netzwerkbasierte Aggregation

Im Rahmen des Projektes IUNO wird ein Konzept zur verteilen netzwerk-basierten Anomalieerkennung entwickelt. In dieser Arbeit wird die dafür notwendige Infrastruktur vorgestellt. Netzwerkbasierte Anomalieerkennung benötigt Netzwerkdaten für die korrekte Funktionsweise. In der Büro-IT ist es meist hinreichend, das Gateway zum Internet zu überwachen, um anormales Verhalten zu erkennen. Da die Gefährdungslage im industriellen Kontext nicht bloß durch Verbindungen nach außen besteht, ist es auch notwendig, innerhalb eines Subnetzes kommunizierende Entitäten zu überwachen. Dies ist durch zentrales Mitlesen des Netzwerkverkehrs nicht möglich. In dieser Arbeit wird daher eine Möglichkeit vorgestellt, durch verteilte Sensoren potentiell jedweden Netzwerkverkehr zu lesen, speichern und bewerten. Da heutzutage auch eingebettete Systeme über nennenswerte Rechenleistung verfügen, ist eine Vorverarbeitung des Datenverkehres auf den lokalen Sensoren möglich. Hier werden die anfallenden Daten mithilfe von Zeitserien-basierten Algorithmen auf Anomalien hin untersucht. Anschließend werden die Daten zu Flows aggregiert, basierend auf einem Aggregationsmodell, und an einer höheren Instanz gesammelt. Dann werden weitere Algorithmen zur Erkennung von Anomalien angewendet und die Daten zu einer zentralen Stelle gesendet. Hier ist ein Überblick über das gesamte Netzwerkverhalten, inklusive der Subnetzwerke, möglich. Um die Daten zu sammeln, wird in jedem Subnetz ein Sensor eingebracht, der die Netzwerkdaten mitschneidet. Diese werden zyklisch über ein Mobile Ad-hoc Network (MANET) an die nächsthöhere Instanz übertragen, nachdem sie aggregiert wurden. Ein MANET wird verwendet, da auf diese Art die Bandbreite des zumeist drahtgebundenen Produktionsnetzwerkes nicht beeinflusst wird.

## 3.1 Problematik

Ein zentrales Element der Angriffserkennung ist die Detektion von Anomalien, sowohl im Netzwerk, als auch auf Host-Systemen. Eine Besonderheit der industriellen Netze liegt in der, im Vergleich zur klassischen Büro-IT, extrem hohen Heterogenität. Dies gilt sowohl für die verwendeten Geräte, als auch für die verwendeten Netzwerkprotokolle. Eine große Herausforderung ist daher die Zusammenführung geeigneter Informationen, um den Sicherheitszustand eines Systems zu beurteilen. Eine weitere Schwierigkeit stellt der Aufbau moderner industrieller Netzwerke dar. So verhindern gemanagte Netzwerke ein zentrales Auslesen aller Verkehrsinformationen an zentraler Stelle. Auch die häufige Unterteilung in Subnetze erschwert die ganzheitliche Datenerfassung an einem zentralen Punkt. Weiterhin gilt es beim Sammeln von Verkehrsdaten stets, die bestehenden Anforderungen an den Kontext, z.B. Sicherheitsniveau oder Latenz nicht zu gefährden. Als in der Industrie allgemeingültige Anforderungen an IT-Security-Systeme haben sich folgende Punkte ergeben:

- Keine Veränderung der Paketlaufzeiten
- Keine Manipulation des Produktivnetzwerks
- Keine Verringerung des Sicherheitsniveaus
- Nachrüstbarkeit für Bestandsmodelle

Im Gegensatz zur klassischen Büro-IT, in deren Kontext sich die Sicherheitsanforderung aus dem klassischen CIA-Modell (Confidentiality – Vertraulichkeit, Integrity – Integrität, Authenticity – Authentizität) ist die höchste Priorität im industriellen Kontext die Verfügbarkeit. Grund hierfür sind die enormen Kosten, die üblicherweise durch Standzeiten anfallen.

## 3.2 Lösungsansatz

Der entwickelte Lösungsansatz erfüllt drei der oben genannten Anforderungen. Durch die Einbringung zusätzlicher Hard- und Software in ein Netzwerk erhöht sich inhärent die Angriffsfläche für einen potentiellen Angreifer. Durch die Verwendung von üblichen Verschlüsselungs- und Authentifizierungsmethoden und weiteren IT-Grundschutz-Standards ist es jedoch ohne großen Mehraufwand möglich, ein bestimmtes Sicherheitsniveau zu gewährleisten. Der Lösungsansatz basiert auf lokalen Kollektoren in den einzelnen Netzsegmenten und globalen Aggregatoren. Dabei muss gewährleistet werden, dass durch die lokalen Kollektoren das jeweilige gesamte Netzsegment observiert werden kann. Eine beispielhafte Topologie ist in Abbildung 2 dargestellt.

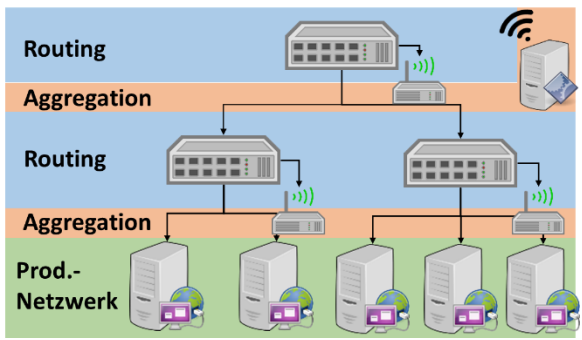

**Abbildung 2: Exemplarische Topologie des entwickelten Konzeptes**

Im dargestellten Beispiel wird der Netzwerkverkehr innerhalb eines Segmentes an den Switches abgegriffen. Dies ist durch technische Maßnahmen, wie etwa Monitoring-Ports, möglich. Der Netzwerkverkehr über Segmentgrenzen hinweg wird an den Gateways abgegriffen. Die in Abbildung 2 dargestellten Router dienen in diesem Szenario sowohl als Router, wie auch als Gateways. Hierbei kann durch die Verwendung von Mirror Ports unter geeigneten Einstellungen sichergestellt werden, dass die Aggregatoren ausschließlich lesend auf das Produktivsystem zugreifen, sodass keine Manipulation desselben möglich ist. Von Interesse ist hierbei der Netzwerkverkehr auf Paketebene, da sich aus diesen Informationen alle relevanten Erkenntnisse gewinnen lassen. Der Erkenntnisgewinn findet lokal auf den Kollektoren statt, lediglich bei Auftreten einer Anomalie wird ein Alarm gemeldet. Als vielversprechende Algorithmen zur Entdeckung von Anomalien haben sich auf dieser Ebene zeitreihenbasierte Verfahren herausgestellt. Darüber hinaus wird von jedem Kollektor ein Flow an den Aggregator gesendet. Dieser Flow enthält zusammengefasste Informationen über den Verkehrsfluss, die im Gegensatz zu den lokal gesammelten Daten ein geringeres Datenvolumen aufweisen. Anhand aller aggregierten Informationen kann der Aggregator das globale Verhalten des Netzwerkes analysieren und bewerten.

Die Kommunikation der Kollektoren mit dem Aggregator geschieht durch ein Mobiles drahtloses Ad-Hoc-Netzwerk (MANET). Dadurch wird die kabelgebundene Infrastruktur der Produktionseinrichtung nicht durch zusätzlichen Verkehr belastet, was zu Veränderungen der Paketlaufzeiten von Produktivdaten führen könnte. Bekannte Vertreter von MANET-Implementierungen sind das Optimized Link State Routing Protocol (OLSR) [26] und better approach to mobile ad-hoc networking (B.A.T.M.A.N.) [27].

Da eine große Anzahl industrieller Netzwerkprotokolle entweder TCP/IP-basiert ist, oder zumindest Ethernet-Schnittstellen verwenden, ist eine einfache Nachrüstbarkeit durch Standard Hard- und Software gewährleistet.

### 3.3 Bewertung

Der beschriebene Lösungsansatz ist geeignet, die in Kapitel 3.1 dargelegten Anforderungen zu erfüllen. Durch die Verwendung eines zusätzlichen Kommunikationskanals wird die Paketlaufzeit des Produktivsystems nicht verändert. Aufgrund des ausschließlich passiven Verhaltens der Kollektoren ist eine Manipulation des Produktivnetzwerkes ausgeschlossen. Durch Anwendung von Methoden des IT-Grundschutzes kann das Sicherheitsniveau zumindest gehalten werden. Bestandsmodelle können durch weitverbreitete Standardschnittstellen mühelos analysiert werden. Weiterhin ist der Ansatz flexibel und skalierbar einsetzbar in dem Sinne, dass eine horizontale Erweiterung durch zusätzliche Kollektoren einfach möglich ist. Eine vertikale Erweiterung ist durch das Hinzufügen weiterer Aggregatoren möglich,3.1 die durch einfache Modifikation hierarchisch angeordnet werden können.

Die Verwendung eines MANET bringt neben vielen Vorteilen auch Nachteile mit sich. Es ist nicht notwendig, a priori eine Infrastruktur bereitzustellen, wie etwa Kabel. MANETs sind aufgrund ihrer inhärenten Eigenschaften robust gegenüber Ausfällen einzelner Teilnehmer und kostengünstig in Pflege und Einsatz [28].

Allerdings ist der Funkkanal per Definition ohne physikalische Zugangsbeschränkung, sodass ein Mithören unautorisierter Entitäten durch technische Maßnahmen (wie etwa Verschlüsselung) verhindert werden muss. Auch der maliziöse Eingriff ist auf diese Art für einen Angreifer möglich, etwa durch das Einbringen von gefälschten Paketen oder durch Jamming. Außerdem kann der Betrieb verschiedener Funknetzwerke in räumlicher Nähe zu Störungen untereinander führen.

## 4  Konklusion und weitere Schritte

In dieser Arbeit wurde das Forschungsprojekt IUNO vorgestellt. Ein besonderer Fokus lag dabei auf am DFKI entwickelten Methoden, die im Rahmen dieses Projektes entstanden sind. Weiterhin wurde ein aktueller Forschungsansatz vorgestellt, der die Problematik der Datensammlung in großen Netzwerken mit dem Ziel der Angriffserkennung behandelt.

Das Forschungsprojekt IUNO adressiert einige der wesentlichen Themen der vierten industriellen Revolution. Deutschland als Land mit historisch starkem Mittelstand ist im Zeitalter der Digitalisierung darauf angewiesen, dass es als Wirtschaftsstandort in der Lage ist, die Sicherheit im Bereich der Informationstechnik zu gewährleisten. Gerade im Produktionsumfeld bestehen hier zurzeit Mängel. Diese werden im Rahmen von IUNO behoben, um skalier- und einfach umsetzbare Lösungen zu entwickeln, die unabhängig von der Unternehmensgröße eingesetzt werden können.

Von Seiten des DFKI wurden mehrere Methoden zum Projekt IUNO beigesteuert, die geeignet sind, die Sicherheit bestimmter Anwendungsfälle zu erhöhen und sich in ein übergreifendes Sicherheitskonzept integrieren lassen.

Ein Schwerpunkt dieser Arbeit lag auf einem System zur Kollektion und Aggregation von Netzwerkdaten in einem Produktionsnetzwerk. Dieses System fügt sich leicht in existierende Netzwerke ein, ohne deren Struktur oder

Verhalten zu beeinflussen. Es löst die Probleme, die sich beim Bedarf nach Daten für Angriffserkennung stellen. Weiterführende Arbeiten bestehen in der Entwicklung und Anpassung ausgefeilter Algorithmen der Anomalieerkennung für die Netzwerkdaten. Auch die Korrelation von Netzwerkdaten mit host-basierten Daten verspricht, die gängigen Lösungen zu verbessern. Weiterhin sollen die Methoden zur Anomalieerkennung mit den Honeypots kombiniert werden, um eine umfassende und automatisierte Analyse des Verhaltens von Schadcode zu erhalten.

# 5 Danksagung

# 6 Literatur